\documentclass[12pt, a4, reqno]{amsart}
\usepackage{hyperref} 
\usepackage{xr}
\usepackage{natbib}
\usepackage{amsmath,amssymb,centernot,mathtools}
\usepackage{amsthm}
\usepackage{graphicx,subcaption}
\usepackage{caption}
\usepackage{enumitem}
\usepackage{color}
\usepackage[table]{xcolor}
\usepackage[calc]{picture}
\usepackage[english]{babel}
\usepackage[margin=1in]{geometry}
\usepackage{secdot}
\usepackage{float}
\usepackage{tikz}
\usepackage{comment}
\usetikzlibrary{arrows.meta,shapes}
\usetikzlibrary{arrows,shapes.arrows,shapes.geometric,shapes.multipart,
decorations.pathmorphing,positioning,shapes.swigs,decorations.pathreplacing}
\usepackage{placeins}
\usepackage{url} 
\usepackage{setspace}
\usepackage{titlesec}
\usepackage{tcolorbox}
\usepackage[outline]{contour}
\usepackage{amsfonts}
\usepackage{algcompatible}
\usepackage{algorithm,algpseudocode,float}
\usepackage{pdflscape}
\usepackage{cancel}
\usepackage{etoolbox}
\usepackage{afterpage}

\renewenvironment{titlepage}{%
  \setcounter{page}{0}
  \thispagestyle{empty}
  \centering
  \vspace*{\fill}
}{%
  \vspace{0\baselineskip}
  \vspace*{\fill}
  \newpage
}
\makeatletter
\patchcmd{\@maketitle}
{\if@titlepage \newpage \else}
{\if@titlepage
 \vspace{\baselineskip}
 \else}
{}{}
\makeatother

\newtheorem{theorem}{Theorem}

\newtheorem{assumption}{Assumption}
\newtheorem{remark}{Remark}

\newtheorem{corollary}{Corollary}
\newcommand\norm[1]{\lVert#1\rVert}

\newcommand\independent{\protect\mathpalette{\protect\independenT}{\perp}}
\def\independenT#1#2{\mathrel{\rlap{$#1#2$}\mkern2mu{#1#2}}}

\DeclareMathOperator*{\bL}{\textit{L}}

\renewcommand\thefootnote{\textcolor{blue}{\arabic{footnote}}}

\makeatletter
\g@addto@macro\normalsize{%
  \setlength\abovedisplayskip{0pt}
  \setlength\belowdisplayskip{0pt}
  \setlength\abovedisplayshortskip{0pt}
  \setlength\belowdisplayshortskip{0pt}
}
\makeatother

\titlespacing*{\section}
  {0pt}{0\baselineskip}{0\baselineskip}
\titlespacing*{\subsection}
  {0pt}{0\baselineskip}{0\baselineskip}
  \titlespacing*{\subsubsection}
  {0pt}{0\baselineskip}{0\baselineskip}

\makeatletter
\newenvironment{breakablealgorithm}
  {
   \begin{center}
     \refstepcounter{algorithm}
     \hrule height.8pt depth0pt \kern2pt
     \renewcommand{\caption}[2][\relax]{
       {\raggedright\textbf{\ALG@name~\thealgorithm} ##2\par}%
       \ifx\relax##1\relax 
         \addcontentsline{loa}{algorithm}{\protect\numberline{\thealgorithm}##2}%
       \else 
         \addcontentsline{loa}{algorithm}{\protect\numberline{\thealgorithm}##1}%
       \fi
       \kern2pt\hrule\kern2pt
     }
  }{
     \kern2pt\hrule\relax
   \end{center}
  }
\makeatother
\begin{document}
\allowdisplaybreaks

\begin{titlepage}
\pagenumbering{gobble}

\title[]{Estimating average treatment effects when treatment data are absent in a target study}

\author[Lan Wen, Aaron L. Sarvet]{Lan Wen$^{\text{a},\ast,\dagger}$, Aaron L. Sarvet$^{\text{b},\ast}$\vspace{0.6cm}
\\ \small
$^\text{a}$Department of Statistics and Actuarial Science, University of Waterloo, Waterloo, Ontario, Canada \vspace{0.3cm} \\
$^\text{b}$Department of Biostatistics \& Epidemiology,
University of Massachusetts, Amherst, USA}
\maketitle



\renewcommand{\thefootnote}{\fnsymbol{footnote}}
\footnotetext[1]{The two authors contributed equally to this paper.}
\footnotetext[2]{corresponding author: lan.wen@uwaterloo.ca}

\vspace{-2em}
\begin{abstract}
Researchers are frequently interested in understanding the causal effect of treatment interventions. However, in some cases, the treatment of interest--readily available in a randomized controlled trial (RCT)--is either not directly measured or entirely unavailable in observational datasets. This challenge has motivated the development of stochastic incremental propensity score interventions which operate on post-treatment exposures affected by the treatment of interest with the aim of approximating the causal effects of the treatment intervention. Yet, a key challenge lies in the fact that the precise distributional shift of these post-treatment exposures induced by the treatment is typically unknown, making it uncertain whether the approximation truly reflects the causal effect of interest. The primary objective of this paper is to explore data integration methodologies to characterize a distribution of post-treatment exposures resulting from the treatment in an external dataset, and to use this information to estimate counterfactual mean outcomes under treatment interventions, in settings where the observational data lack treatment information and the external data \textit{may} not contain measurements of the outcome of interest.
We will discuss the underlying assumptions required for this approach and provide methodological guidance on estimation strategies to address these challenges.

\end{abstract}
\sloppy
\end{titlepage}

\pagenumbering{arabic}  
\section{Introduction}
To make an informed decision about a novel treatment, patients often rely on evidence about its long-term treatment effects (LTEs). While randomized controlled trials (RCTs) are the gold standard for estimating treatment effects, they may not capture long-term outcomes. In contrast, observational studies routinely measure long-term outcomes, but are limited by the lack of randomization.
In many cases, the treatment variable of interest is either not directly measured or a particular level of the treatment (e.g., the new medication) is entirely unavailable to units in the observational data. 

In the face of these challenges, investigators have considered the effects of so-called \textit{stochastic interventions}, which we henceforth refer to as stochastic exposure effects (SEEs). Formally, SEEs are defined by interventions which assign some measured ``exposure" (distinct from the original treatment of interest) randomly according to a pre-specified distribution \citep{Munoz2012,Kennedy2019,Young2019,diaz2020non,Wen2021,sewak2024causal}. 
These estimands would seemingly have little policy relevance, as a decision-maker in the real-world (outside of an experimental setting) would rarely consider taking any treatment according to a random assignment rule. However, these SEEs for the measured exposure are often interpreted instead as representing the LTE of the original unmeasured treatment of interest -- an interpretation that is not explicitly justified.

In this paper, we formalize the link between LTEs and SEEs. Specifically, we formally consider the assumptions necessary for their values to coincide, which would thereby justify interchanging interpretations of any results. A key assumption is that the SEE's pre-specified exposure distribution is equivalent to that induced by intervening on the treatment of interest. We argue that this assumption is often implausible and lacks empirical justification. Rather than positing a hypothetical distribution for the post-treatment exposures, we propose a data integration approach. The resulting methods leverage data from an external trial for the treatment on the exposure measured in the observational data.

\subsection{Related literature.}
~Multi-source data is becoming increasingly common \citep{olsen2007learning, fiore2016integrating}. 
The integration of these data can improve accuracy, which is a core objective of traditional meta-analysis \citep{schmid2020handbook}.
For instance, integrating RCT with observational data can improve the statistical power of treatment effect estimates \citep{lin2024data}.
By integrating data from different sources, researchers can also extend causal inferences to a broader target population \citep{dahabreh2020extending,dahabreh2020toward, dahabreh2023efficient, colnet2024causal}. A growing body of methodological research develops this approach \citep{cole2010generalizing, Pearl2014, rudolph2017robust, westreich2017transportability, dahabreh2020toward, guo2021multi, colnet2024causal}.

Data integration has also been leveraged for the general objective of learning treatment effects on long-term outcomes, when a single data source is insufficient 
\citep{rudolph2017robust,athey2020combining,cheng2021adaptive, cheng2022long,ghassami2022combining,robbins2024data, imbens2024long,kallus2025role,chen2025nonparametric}. 
For example, 
\cite{rudolph2017robust} studied methods for transporting intention-to-treat effects from one population with long-term follow-up outcome data to another population without.
\cite{athey2020combining} target a binary treatment effect on an outcome observed only in the observational sample (see also \citealp{imbens2024long}). In the same spirit, \cite{robbins2024data} leverage short-term surrogate outcomes (observed in both datasets) to multiply impute missing long-term outcomes in the experimental data, addressing a setting where treatment is not measured in the observational study. Collectively these methods suppose that the only missing element is the long-term outcome in the experimental data. We consider a setting where additionally the treatment is unmeasured in the non-experimental data. This setting was independently considered in \cite{athey2024surrogate}. Our current work is distinguished in at least four different ways. First, we explicitly consider a related setting where the treatment variable is measured in the non-experimental data, but there is zero-probability of observing a value of interest for that treatment variable (e.g., a novel medication or policy). 
Second, we make a direct theoretical comparison between assumptions identifying SEEs and LTEs, thereby bridging the two frameworks and clarifying their relationship.
Third, we address both cases in which the outcome of interest is only observed in the target source and cases in which it in observed in both target and external sources.
Finally, we propose a novel sample-bounded semiparametric estimator derived from the efficient influence function and establish its unique asymptotic properties, including: (1) triple robustness under working models for the nuisance functions, and (2) an explicit expansion of the first-order bias term, which reveals the convergence rate conditions necessary for the nuisance estimators.

\subsection{Motivating examples.}
~A motivation for studying SEEs is that in many instances, the treatment interventions being studied in a trial are not measured or not yet available in the broader population (see Web Appendix A).
This precludes us from understanding the effect of these new treatments on some outcome of interest in observational studies.
To motivate the settings we study, we provide two illustrative examples below, with two additional examples on cancer screening and gene therapy provided in Web Appendix C.

\subsubsection*{Example 1. }
~Suppose we are interested in evaluating the effect of new smoking cessation interventions (``treatment'' $A$), measured only in a trial, on long-term outcomes such as five-year all-cause mortality (``outcome'' $Y$), available only in an observational data. To enable this, we leverage post-treatment smoking status (post-treatment/mediating ``exposure'' $M$), which is measured in both data sources and acts as a mediating link between $A$ and $Y$. 
A detailed illustration of this example is provided in Section \ref{sec:datanalysis}. 

\subsubsection*{Example 2. }
~Consider studies on initiation of Pre-Exposure Prophylaxis (PrEP).
Previous research has consistently found that certain high-risk groups have very low rates of starting PrEP, despite their increased risk of HIV infection. 
A realistic intervention would be one aimed at increasing PrEP uptake within such populations or subpopulations.
For instance, \cite{chan2021randomized} consider behavioral interventions ($A$) intended to enhance PrEP awareness and utilization ($M$) among high-risk individuals. 
These behavioral interventions are often not measured in observational studies (e.g., \citealp{Wen2021,sewak2024causal}), and the trial by \cite{chan2021randomized} does not evaluate their long-term effect on HIV incidence ($Y$). 


\subsection{Organization of the paper.}
~In Section \ref{sec:datastructure}, we introduce notation, the data structure, and our causal estimand. In Sections \ref{sec:design}--\ref{sec:estimation}, we describe identification results and a new sample-bounded semiparametric estimator.
In Section \ref{sec:simulations}, we present results from a simulation study, and in Section \ref{sec:datanalysis}, we present a study transporting the treatment effect of smoking cessation interventions to data from the National Health and Nutrition Examination Survey. 

\section{Data structure}
\label{sec:datastructure}
\subsection{Notations.}
~Suppose that we have random samples from an external source population ($S=1$; e.g., observed via an external source such as an RCT study) and from a target population of interest ($S=0$; e.g., observed via a target source such as an observational study).
Let $A$ denote a discrete treatment variable that would define a LTE, $M$ an observed post-treatment exposure variables (e.g.,~the primary outcome in the RCT) that would define an SEE, $L$ a vector of covariates measured at baseline, and $Y$ a long-term outcome of interest; for any random variable $X$ we denote its support by $\mathcal{X}$ across both sources.
For notational simplicity, we assume throughout that all covariates are discrete in that they have distributions that are absolutely continuous with respect to a counting measure but arguments naturally extend to settings with continuous covariates and Lebesgue measure. 

For each observation $i$, we observe the following data: $O_i=(S_i,\bL_i,A_i^*, M_i, Y_i^*)$, where $A_i^*\in \{\mathcal{A},\emptyset\}$ and  $Y_i^*\in \{\mathcal{Y},\emptyset\}$ represent partial measurements of $A_i$ and $Y_i$, respectively, and are linked by assumptions. 
In one setting we assume that all levels of treatment are administered in both sources but only observed in the external source, while the long-term outcome $Y$ is only observed the in target source (see Table \ref{tab:data_avail}). 
We also consider a setting where the treatment is fully measured in both populations, but the treatment level of interest is unavailable in the target population.
Let $n=n_{s=1}+n_{s=0}$ denote the total number of observations from both studies.
Henceforth, we omit subscripts $i$ unless needed. 

\begin{remark} \label{remark: recant}
Suppose an investigator initially postulates an exposure variable $\tilde M$. We defined $L$ as baseline covariates, that occur temporally prior to treatment $A$. These variables, could in principle cause any combination of $(A, \tilde M, Y)$. However, in many settings it is essential to also consider post-treatment variables that could commonly cause $\tilde M,$ and $Y$, and may themselves be cause by $A$. Let $L_1$ denote the pre-treatment covariates and $L_2$ denote the post-treatment covariates. In the course of specifying a causal model, an investigator might assume that some subset of $L_2$ is not caused by treatment. Let $L_2'$ denote this subset and $L_2''\equiv L_2\setminus L_2'$ be the remaining post-treatment covariates that may be caused by treatment. A covariate in $L_2''$ is sometimes referred to as exposure-induced mediator-outcome confounder or a recanting witness. For the purposes of the following results, we define $M\coloneqq \{\tilde M, L_2''\}$ and $L \coloneqq \{L_1, L_2'\}$. In other words, all recanting witness covariates are considered ``exposures'' and all covariates are considered considered to be topologically pre-treatment.  Further details are provided in Web Appendix G, where we explain this using PrEP Example 2.
\end{remark}

\begin{table}
    \centering
    \begin{tabular}{l|c}
       Data from population & Available data \\ \hline
       $S=0$: all levels of $A$ available but unmeasured  & $L,M,Y$ \\
         $S=0$: only placebo or standard treatment available  & $L,A=0,M,Y$ \\
       $S=1$: outcomes unmeasured & $L,A,M$ \\
       \hline
    \end{tabular}
    \caption{Data availability for the observational study (from $S=0$) and the randomized controlled trial (from $S=1$). $L$ denotes the set of baseline covariates, $A$ the treatment variable, $M$ the exposure variable, and $Y$ the outcome.}
    \label{tab:data_avail}
\end{table}

\subsection{Target Estimand.}
~Let $Y^{a}$ denote the counterfactual (potential) outcome variable if, possibly contrary to fact, the treatment {were set to} a value $a$ for $a\in \mathcal{A}$. 
Our aim is to estimate the estimand defining an LTE, given by $\Psi^a \coloneqq \mathbb{E}(Y^{a}\mid S=0)$, which denotes the average counterfactual long-term outcome had treatment been set to some value $a$ in a target population. 
We also consider the estimand defining an SEE, given by $\Psi^g \coloneqq \mathbb{E}(Y^{g}\mid S=0)$, which denotes the average counterfactual long-term outcome under an intervention $g$ that randomly assigns levels of the post-treatment exposure $m\in\mathcal{M}$ based on a pre-specified conditional distribution $p^g(m \mid l)$, where assignment depends on covariates $L=l$.
 An investigator who motivates an LTE but explicitly targets $\Psi^g$ implicitly assumes that $\Psi^g = \Psi^a$. 
Alternatively, an investigator could target $\Psi^a$ without pre-specifying $p^g(\cdot)$, and instead learn this distribution for $M$ from data.
Our assumptions enable us to transport treatment effects on the exposure from the RCT to the target population of interest, thereby achieving identification of $\Psi^a$.

\section{On study design}
\label{sec:design}

Consider two sampling models: (1) a population sampling model where data are obtained by simple random sampling from a common super-population; (2) a biased sampling model where the
sampling is done stratified on $S$.
We use $p(\cdot)$ to denote densities from the population sampling model, and $q(\cdot)$ to denote densities from the biased sampling model, with respect to a suitable dominating measure. 
Under a population sampling scheme, the density of the full factual data is given by: $p(s,l,a,m,y)=p(s)p(l\mid s)p(a\mid l,s)p(m\mid a,l,s)p(y\mid m,a,l,s)$; under a biased sampling scheme, the full data density would be: $q(s,l,a,m,y)=q(s)p(l\mid s)p(a\mid l,s)p(m\mid a,l,s)p(y\mid m,a,l,s)$, where $q(s) \coloneqq P_q(S=s)$, and most generally $q(s) \neq p(s)$. 
See \cite{kennedy2015semiparametric} for a similar argument in the context of matched cohort studies and \cite{dahabreh2020extending,dahabreh2021study,dahabreh2023efficient} for analogous arguments in transportability analyses.
We adopt the biased sampling model framework throughout.

\section{Identification: Trial Treatments Are Available in Real World}
\label{sec:identification}
We begin with the setting in which all treatment levels of interest exist in both study but are unobserved in $S=0$. 
In Section \ref{sec:special}, we describe a distinct setting where a treatment level is not yet accessible outside the trial, as is often the case.

\subsection{Identifying assumptions.}
~To express conditions on distributions under hypothetical interventions on the treatment variable $A$, we adopt the convention $W^a \independent_d X^a \mid Z^a$ to denote that the conditional distribution of $W^a$ does not depend on $X^a$, given any value of $Z^a$, where $\independent_d$ indicates d-separation. 
Following \citet{richardson2023potential}, we write  $W^a \independent_d~ a \mid Z^a$ to denote that the conditional distribution of $W^a$ given any value of ($Z^a,A$) does not depend on $a$.\footnote{That is, $p(W^a\mid Z^a, A)$ does not depend on $a$.} These conditions can be directly interrogated using Single World Intervention Graphs (SWIGs; \citealp{richardson2013single}), where d-separation implies the corresponding functional independence. 

\begin{figure}
\begin{minipage}{1\linewidth}
\centering
 \begin{tikzpicture}
                    \tikzset{line width=1pt, outer sep=0pt,
                    ell/.style={draw,fill=white, inner sep=3pt,
                    line width=1pt},
                    swig vsplit={gap=5pt,
                    inner line width right=0.5pt}};
\begin{scope}[every node/.style={thick,draw=none}]
    \node[name=S,ell,  shape=ellipse] at (-3,0) {$S$};

    \node[name=A,shape=swig vsplit] at (1.25,0) {
                    \nodepart{left}{$A$}
                    \nodepart{right}{$a$}   };
    \node[name=M,ell,  shape=ellipse] at (6,0) {$M^{a}$};
    \node[name=Y,ell,  shape=ellipse] at (10,0) {$Y^{a}$};	
    \node[name=U1,ell,  shape=ellipse] at (-1,1.5) {$U$};
    \node[name=U2,ell,  shape=ellipse] at (2,3.25) {$U$};
    \node[name=U3,ell,  shape=ellipse] at (1,2.5) {$U$};
     \node[name=U4,ell,  shape=ellipse] at (4.5,1.5) {$U$};
    \node[name=U5,ell,  shape=ellipse] at (2.75,1.5) {$U$};

\end{scope}
\begin{scope}[>={Stealth[black]},
              every node/.style={fill=white,circle},
              every edge/.style={draw=black,thick}]
    \path [->] (U1) edge[color=red, very thick] (A);
    \path [->] (U1) edge[color=red, very thick]  (S);
        \path [->] (U2) edge[bend right=30]  (S);
        \path [->] (U2) edge[bend left=10]  (Y);

    \path [->] (U3) edge[color=red, very thick]  (A.120);
    \path [->] (U3) edge[bend left=10, color=red, very thick]  (Y);
    \path [->] (U4) edge[color=red, very thick]  (M);
    \path [->] (U4) edge[color=red, very thick]  (Y);
    \path [->] (U5) edge[color=blue!60, very thick]  (A.120);
    \path [->] (U5) edge[color=blue!60, very thick] (M.170);

    \path [->] (S) edge (A);
    \path [->] (S) edge[bend right]  (Y);
    \path [->] (A) edge  (M);
    \path [->] (A) edge[bend right, color=red, very thick]   (Y);
    \path [->] (M) edge (Y);

\end{scope}
\end{tikzpicture}
\end{minipage}
\caption{SWIG compatible with the identifying assumptions. The distribution in $S=1$ is governed by the sub-SWIG that \textit{excludes} the blue arrows; the distribution in $S=0$ is governed by the sub-SWIG that \textit{excludes} the red arrows. $U$ denotes unmeasured variables.}
\label{fig:dataintcomplex}
\end{figure}
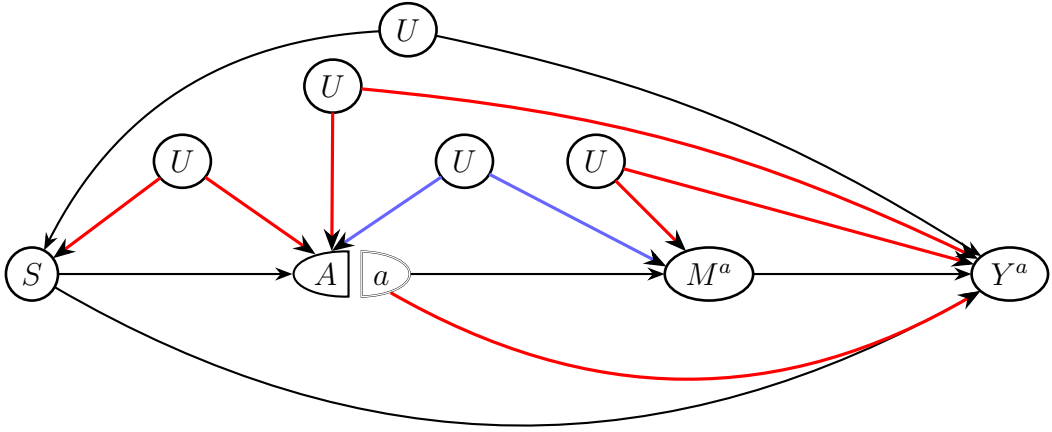

Some of the conditions we consider are also crucially context-dependent, that is they specifically concern either the subgroup defined by the sample from the external source population ($S=1$) or the target population ($S=0$). To encode these context-dependencies on a single SWIG, we color-code the arrows: the distribution in the trial is governed by the sub-SWIG that \textit{excludes} the blue arrows; the distribution in the target source is governed by the sub-SWIG that \textit{excludes} the red arrows. We give an example of such a SWIG in Figure \ref{fig:dataintcomplex}, where we have also excluded measured covariates $L$ to avoid clutter. An investigator could construct such a SWIG by only including a red arrow between two nodes if a direct causal effect was operative in the trial but not the target source, a blue arrow if an effect was operative in the target source but not the trial, and a black arrow if it was operative in both.

Given this, consider the following set of assumptions. The first assumption (Assumption \ref{ass:consist}) requires that the intervention on treatment is well-defined (i.e., no interference and no multiple versions of treatment; \citealp{vanderweele2013causal}).

\allowdisplaybreaks
\begin{assumption}[Consistency for treatment]
If $A=a$, then $M^a=M$ and $Y^a=Y$, $\forall a\in \mathcal{A}$;
    \label{ass:consist}
\end{assumption}

  \begin{assumption}[No unmediated treatment confounding or direct effects for $Y$ in $S=0$]
  $$Y^{a}\independent_d (a, A) \mid (M^a, L,S=0), ~~\forall a\in \mathcal{A};$$
    \label{ass:zyexch1a} 
 \end{assumption}
 \vspace{-2em}
 
 \begin{assumption}[No treatment confounding for the exposure in $S=1$]
  $$M^a\independent_d A \mid (L,S=1),~~\forall a\in \mathcal{A};$$
    \label{ass:zyexch1b} 
 \end{assumption}
  \vspace{-2em}

   \begin{assumption}[Source irrelevance for the exposure]
    $$M^a\independent_d S\mid L,~~\forall a\in \mathcal{A}.$$
    \label{ass:treatsource1}
\end{assumption} 
\vspace{-2em}

Let model $\mathcal{M}^{di}$ denote the \textit{data integration} model, i.e., the set of laws satisfying Assumptions \ref{ass:zyexch1a}--\ref{ass:treatsource1}. These assumptions hold on the SWIG of Figure \ref{fig:dataintcomplex}.  Assumption \ref{ass:zyexch1a} encodes both treatment exclusion restriction \textit{and} $A-Y$ conditional exchangeability in $S=0$. It would be contradicted by a direct effect of treatment on the outcome ($a\rightarrow Y^{a}$), unmeasured treatment-outcome confounding ($A\leftrightarrow Y^{a}$), M-bias due to the exposure ( $A\leftrightarrow M^a \leftrightarrow Y^{a}$), collider bias due to the exposure ($A\rightarrow M^a \leftrightarrow Y^{a}$), or M-bias due to the source population ($A\leftrightarrow S \leftrightarrow Y^{a}$).  
Assumption \ref{ass:zyexch1b} concerns independencies in $S=1$. It would be contradicted by unmeasured treatment-exposure confounding ($A\leftrightarrow M^{a}$; e.g., due to non-compliance) or M-bias due to data source ($A\leftrightarrow S \leftrightarrow M^{a}$). 
Finally, Assumption \ref{ass:treatsource1} links the trial and the target source. It would be contradicted by any unmeasured common causes of the source and the exposure ($S\leftrightarrow M^{a}$), or any direct effects of the source on the exposure ($S\rightarrow M^{a}$). 

\begin{remark}
The context-specificity of Assumptions \ref{ass:zyexch1a}--\ref{ass:zyexch1b} is important because the assumptions may be plausible even when their context-independent analogues fail.
For example, certain trials offer post-follow-up benefits that augment treatment and directly affect $Y$ (e.g., the Look AHEAD clinical trial, \citealp{rushing2017cost}; the Optimune trial, \citealp{holtdirk2021results}); these trial-specific features justify a direct arrow from context to the outcome, $S \rightarrow Y^a$, or an arrow from treatment to the outcome, $a \rightarrow Y^a$, in the SWIG. This aligns with our context-specific assumptions but not their context-independent counterparts. Also, the determinants of treatment in a trial (i.e., randomization, non-compliance factors) may differ from those in the target population; specifically, trials may avoid short-term $M-A$ unmeasured confounding, thereby justifying the absence of $A \leftrightarrow M^{a}$ in the SWIG. 
These context-specific effects of treatment and confounding complicate LTE identification using either data source alone\footnote{Even in the setting where $(L,A,M,Y)$ are observed in both sources (see SWIG in Figure \ref{fig:dataintcomplex}).} but remain compatible with our data integration approach for the LTE in the target source.
\end{remark}

\begin{remark}
Assumptions \ref{ass:consist}--\ref{ass:zyexch1a} jointly imply that the conditional distribution of the outcome given the exposure and covariates is not a function of the treatment in $S=0$, i.e., $Y\independent A\mid (M,L,S=0)$. As such, if data on $A$ were available in $S=0$, this distributional implication could be empirically tested and potentially falsified.  
\end{remark}

\begin{assumption}[Positivity]
~~
\begin{enumerate}[label=(\roman*).]
    \item If $p(l\mid S=0)>0$ then $p(l\mid S=1) > 0~\wedge~p(a\mid l, S=1) > 0$ $\forall l$;
    \item If $p(l\mid S=0)>0~\wedge~p(m \mid l, a, S=1) > 0$ then $p(m\mid l, S=0)>0$ $\forall  l,m$.
\end{enumerate}

    \label{ass:positiveA}
\end{assumption}

The first positivity condition asserts that all treatment level $a$ must be observable in the trial for all covariate patterns that are observable in the target population. 
The second positivity condition asserts that, within these same covariate patterns, we must observe all levels of the exposure in the target source population that would arise in the corresponding strata of the trial that took treatment level $a$. Under Assumptions \ref{ass:zyexch1b}--\ref{ass:treatsource1}, these are the levels of the exposure that would arise under an intervention that sets treatment to level $a$.  

\begin{assumption}[Partially observed treatments and outcomes]
~~
\begin{enumerate}[label=(\roman*).]
\item $P(A^*=A \mid S=1)=1 ~\wedge~ P(A^\ast=\emptyset \mid S=0)=1$;
\item $P(Y^*=Y \mid S=0)=1~\wedge~P(Y^*=\emptyset \mid S=1)=1$.
\end{enumerate}
    \label{ass:partial}
\end{assumption}

Assumption \ref{ass:partial} simply states that the outcome is measured only in the target population and the treatment is only measured in the trial population. 

\begin{theorem}
Under Assumptions \ref{ass:consist}--\ref{ass:partial}, the average counterfactual outcome had treatment been set to a value $a$ for $a\in \mathcal{A}$ in data source $S=0$, i.e., $\Psi^a \coloneqq E(Y^{a}\mid S=0)$, is identified by the following g-formula:
\begin{align}
    \Psi^a & = \smash\sum_{m,l}E(Y\mid M=m,L=l,S=0) 
        P(M=m\mid L=l,A=a, S=1)P(L=l\mid S=0).
        \label{eq:identeq1lte}
\end{align}
    \label{thm:ident1}
\end{theorem}
\vspace{-2em}

Theorem \ref{thm:ident1} illustrates how we can identify the average counterfactual outcome under interventions on treatment $A$ via data integration of trial and observational data. In Web Appendix B, we present an alternative set of sufficient assumptions for identification that resembles a close connection to the assumptions for the frontdoor algorithm \citep{pearl1993mediating, pearl2009causality,Fulcher2020,wen2024causal}.
In Web Appendix G, we extend the data integration framework to the PrEP Example 2, where the treatment intervention follows a covariate-dependent (i.e., dynamic) strategy based on $L$.


\begin{corollary}
Suppose $Y$ is measured for all subjects in $S=1$. That is, we consider Assumption \ref{ass:partial}* which retains part (i) of Assumption \ref{ass:partial} and replaces part (ii) with the condition $P(Y=Y^*)$. Under this modified assumption, impose that $Y\independent_d S\mid (M,L)$ leads to a different identification result. 
Specifically, Expression \eqref{eq:identeq1lte} simplifies to:
\begin{align}
    \Psi^a= \smash\sum_{m,l}E(Y\mid M=m,L=l) 
        P(M=m\mid L=l,A=a, S=1)P(L=l\mid S=0)
        \label{eq:identeq-strong}
\end{align}
which allows one to leverage information across data sources to model the outcome, thereby improving efficiency (see Section \ref{sec:special} for further discussion).
\end{corollary}

\subsection{Interpretation of stochastic exposure effects as long-term effects.} \label{sec: LTE-SEE}
~As before, let $g$ denote the regime that randomly assigns the exposure according to a pre-specified distribution $p^g(m\mid l)$ for an SEE. 
To facilitate discussion in this section, we suppose treatments, exposure, and outcome are binary. As part of our regularity conditions, we assume that $M$ and $L$ are defined such that, graphically, there are no exposure-outcome confounders caused by treatment (see Remark \ref{remark: recant}). We further assume that the investigator regards treatment as a cause of the exposure in the population; that is, $p^{a}(m \mid l) \coloneqq P(M^{a}=m \mid L=l, S=0)$ differs from $p(m \mid l)$ for at least one $l$. Finally, to avoid uninteresting technical subtleties, we assume the law of $(S,L,A,M)$ is strictly positive.  

To identify an SEE, investigators typically assume that all exposure-outcome confounders are measured in $L$ and that interventions on the exposure is well-defined. Formally, they consider a model defined by the following assumptions.

  \begin{assumption}[Backdoor model for stochastic exposure effects]
  ~~
\begin{enumerate}[label=(\roman*)]
   \item $Y^{m}\independent_d M \mid (L,S=0) 
   ~\forall ~m;$ 
\end{enumerate}
    \label{ass: BD}
 \end{assumption}
Under Assumption \ref{ass: BD} (see \citealp{Kennedy2019}) and the regularity conditions stated above (see Web Appendix B), an investigator can then identify $\Psi^g$ via the g-formula:
\begin{align}
    \Psi^g &\coloneqq E(Y^{g}\mid S=0) \nonumber
    \\ & = \smash\sum_{m,l}E(Y\mid M=m,L=l,S=0) 
        p^g(m \mid l, S=0)P(L=l\mid S=0). \label{eq: SEEgform}
\end{align}
However, SEEs are rarely interpreted as such. \citet{diaz2013assessing} describe a stochastic intervention parameter as ``\textit{the causal effect of a realistic intervention [on a treatment] that intends to alter the population distribution of a [post-treatment] exposure}''. As an example, they considered ``\textit{the use of mass media messages [the treatment, $A$] advertising condom use [the exposure, $M$] as a means of prevention of HIV infection [the outcome, $Y$]}'' \citep[pg. ~161]{diaz2013assessing}; an investigator targeting an SEE would pre-specify a counterfactual distribution of condom use ($M$) under mass media messaging via $p^g(\cdot)$.

Accordingly, there is an implicit assumption that interventions on the pre-exposure treatment are well-defined (Assumption \ref{ass:consist}) and that the pre-specified exposure distribution coincides with its counterfactual distribution under an intervention setting treatment to $a$. In this section, we make the assumptions of the SEE strategy for identifying the LTE explicit and contrast them with the assumptions of the data integration strategy proposed in Section \ref{sec:identification}. Formally, the SEE strategy requires the following augmenting assumption:

\begin{assumption}[Oracle exposure law]
 $$p^g(m \mid l) = p^{a}(m \mid l) \coloneqq P(M^{a}=m \mid L=l, S=0), ~~ \forall \ l, m.$$ \label{ass: oracle}
\end{assumption}
\vspace{-2em}

Let $\mathcal{M}^{see}$ denote the model defined by Assumptions \ref{ass:consist}, \ref{ass: BD}, and \ref{ass: oracle}, together with the regularity conditions stated at the beginning of this section. Although not immediately apparent, it follows that $\Psi^{g}=\Psi^{a}$ under $\mathcal{M}^{see}$; a formal proof is given in Web Appendix B. 
Heuristically, Assumption \ref{ass: BD}(i) precludes $A$ from being a confounder of $M$ and $Y$. 
Since $A$ is a cause of $M$, this precludes any unmeasured common causes of $A$ and $Y$ or any direct effect of $A$ on $Y$ not mediated by $M$.
Combined with Assumption \ref{ass: BD}(ii), it follows that $E(Y\mid M=m,L=l,S=0) = E(Y^a\mid M^a=m,L=l,S=0)$. Substituting this into Expression \eqref{eq: SEEgform}, and noting that $p^g(\cdot)=p^a(\cdot)$ via Assumption \ref{ass: oracle}, it follows by the laws of probability that $\Psi^{g}=\Psi^{a}$. 

\begin{figure}
\begin{center}
\begin{tikzpicture}
  \def\r{2}     
  \def\A{-2}    
  \def\B{0}     

  \begin{scope}
    \clip (\A,0) circle (\r);
    \clip (\B,0) circle (\r);
    \fill[red!50] (-5,-2.5) rectangle (3,2.5);
  \end{scope}

  \begin{scope}
    \clip (\B,0) circle (\r);
    \begin{scope}[clip,even odd rule]
      \clip (\B,0) circle (\r)
            (\A,0) circle (\r);
      \fill[purple!20] (-1,-2.5) rectangle (3,2.5);
    \end{scope}
  \end{scope}

  \begin{scope}
    \clip (\A,0) circle (\r);
    \begin{scope}[clip,even odd rule]
      \clip (\A,0) circle (\r)
            (\B,0) circle (\r);
      \fill[green!40] (-5,-2.5) rectangle (0,2.5);
    \end{scope}
  \end{scope}

  \draw[draw=black, line width=2pt, fill=none] (\A,0) circle (\r);
  \draw[draw=black, line width=2pt, fill=none] (\B,0) circle (\r);

  \draw[thick] (\A,0) -- (\B,0);
  \draw[thick] (\B,0) -- ({\B + \r*cos(45)}, {\r*sin(45)});

  \node at (-1,0.5) {\footnotesize $BD1$};
  \node at (0,1.5) {\footnotesize $BD2$};
  \node at (1,-0) {$Oracle$};
  \node at (-1,-1) {$IV$};
  \node at (-3,0) {$Transp.$};

  \draw [decorate,decoration={brace,amplitude=10pt}, thick]
    (-2.3,-2.7) -- (-4.8,0) node[midway, sloped, below=10pt, anchor=east, rotate=45] {\footnotesize $\mathcal{M}^{di}$};

  \draw [decorate,decoration={brace,mirror,amplitude=10pt}, thick]
    (0.3,-2.7) -- (2.8,0) node[midway, sloped, below=10pt, anchor=west, rotate=-45] {\footnotesize $\mathcal{M}^{see}$};

  \begin{scope}[shift={(3.5,1.5)}]
    \draw[fill=green!40,draw=black] (0,0) rectangle +(0.4,0.4);
    \node[right=5pt] at (0.4,0.2) {\footnotesize $\mathcal{M}^{transport}$};

    \draw[fill=red!50,draw=black] (0,-0.6) rectangle +(0.4,0.4);
    \node[right=5pt] at (0.4,-0.4) {\footnotesize $\mathcal{M}^{bd-IV}$};

    \draw[fill=purple!20,draw=black] (0,-1.2) rectangle +(0.4,0.4);
    \node[right=5pt] at (0.4,-1.0) {\footnotesize $\mathcal{M}^{oracle+}$};
  \end{scope}
\end{tikzpicture}
\end{center}
\caption{Diagram depicting relations between assumptions for identifying long-term effects (LTE) via a data integration (DI) approach and a stochastic exposure effect (SEE) approach. $BD1$ denotes the absence of any backdoor paths between $M$ and $Y$ conditional on $L$. $BD2$ denotes $M$-consistency (Assumption \ref{ass: BD}(ii)) used to identify the SEE. $IV$ denotes instrumental variable independence conditions (under which Balke-Pearl bounds are sharp; see Web Appendix B) and $A$-consistency (Assumption \ref{ass:consist}). 
$IV$ and $BD1$ considered jointly comprise $\mathcal{M}^{bd-IV}$. 
$Transp.$ indicates the exchangeability assumption used to identify the effect of the treatment on the exposure in the external source population ($S=1$), and the assumption that would allow transportability of that effect (Assumptions \ref{ass:treatsource1} and \ref{ass:zyexch1b}, respectively); together they comprise the sub-model $\mathcal{M}^{transport}$. $Oracle$ represents oracle knowledge of the effect of the treatment on the exposure (Assumption \ref{ass: oracle}), which together with $BD2$ comprise the sub-model $\mathcal{M}^{oracle+}$. Under our regularity conditions, the assumptions $(IV,~ BD1,~ Transp.)$ define the DI approach model $\mathcal{M}^{di}$, and $(IV, ~BD1, ~BD2,~ Oracle)$ define the SEE model $\mathcal{M}^{see}$.} \label{fig: SEE-DI comp}
\end{figure}
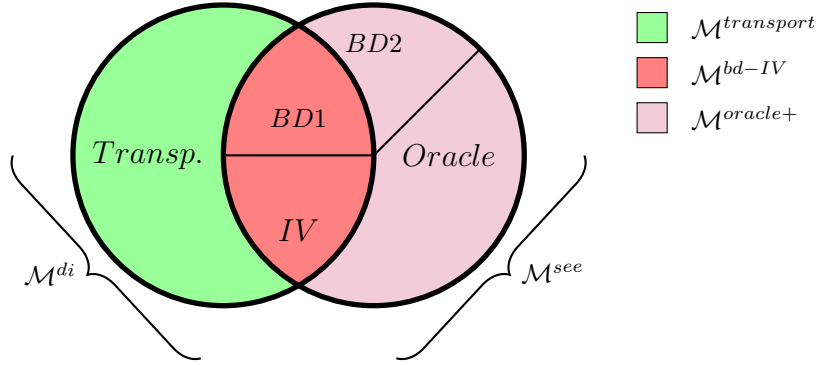

Thus, models $\mathcal{M}^{see}$ and $\mathcal{M}^{di}$ provide two strategies for identifying the LTE. We depict the relationship between $\mathcal{M}^{see}$ and $\mathcal{M}^{di}$ in Figure \ref{fig: SEE-DI comp}. To summarize, let $\mathcal{M}^{oracle+}$ denote the set of laws satisfying Assumption \ref{ass: oracle} and the exposure consistency condition (Assumption \eqref{ass: BD}(ii)); let $\mathcal{M}^{transport}$ be the set of laws that permit identification of $p^{a}(m \mid l)$ from the external source; and let $\mathcal{M}^{bd-IV}$ be the set of laws that both satisfy the backdoor graphical conditions for the SEE, and classical instrumental variable 
conditions for the LTE (see Web Appendix B).  
\begin{theorem}
With the models defined above, the following decompositions hold: $\mathcal{M}^{see} \equiv \mathcal{M}^{oracle+} \cap \mathcal{M}^{bd-IV}$ and $\mathcal{M}^{di} \equiv \mathcal{M}^{transport} \cap \mathcal{M}^{bd-IV}.$
    \label{thm:LTE SEE theorem}
\end{theorem}
This decomposition clarifies the relation between $\mathcal{M}^{see}$ and $\mathcal{M}^{di}$ and illustrates their relative trade-offs. An investigator who rejects $\mathcal{M}^{see}$ in favor of $\mathcal{M}^{di}$ trades assumptions about consistency of a post-treatment exposure and oracle knowledge of its counterfactual distribution (\ref{ass: BD}(ii) and \ref{ass: oracle}) for assumptions that allows this distribution to be identified and transported from an external source (\ref{ass:zyexch1b} and \ref{ass:treatsource1}). 
In many cases $\mathcal{M}^{di}$ will be preferred, because: 1) oracle knowledge of counterfactual distributions is rare; 2) interventions on exposures are often ill-defined; and 3) investigators increasingly have access to clinical trials drawn from populations exchangeable with the target population, as in pragmatic clinical trials.

\section{Trial treatment unavailable in the target population}
\label{sec:special}
In many settings, a treatment variable may be measured in the target population, but a specific treatment level of interest, e.g. a new experimental treatment only available to trial participants, has probability zero in $S=0$. 
In this case, Assumption \ref{ass:partial} is violated.
Investigators may instead consider a modification of Assumption \ref{ass:partial} that requires (i) $P(A^*=A)=1$ but retains condition (ii) as stated. 
Under this modified condition, we consider an alternative estimand motivated by recent developments in separable effects \citep{stensrud2021generalized, stensrud2022separable, robins2022interventionist}, which likewise involve estimands defined for treatment levels with zero probability in the target population \citep{pitts2025addressing}. 
We consider this estimand to ensure that, in $S=0$ where a new treatment is not yet accessible, there are no unmeasured common causes of treatment and outcome once the treatment does become accessible.

We introduce a population-level variable $V$ to indicate the accessibility of a new treatment under study within a data source where $V = 1$ indicates that a new treatment ($A=1$, or multiple new treatments, e.g., $A=1$ and $A=2$) is accessible within that source population, possibly through a particular means, and $V = 0$ indicates otherwise. 
As before, $S = 1$ corresponds to individuals in an RCT, and $S = 0$ corresponds to nonparticipants from a target population. We assume that those in $S=1$ can receive both $A=1$ and $A=0$ (or $A=1$ and possibly another type of novel treatment, say $A=2$), while those in $S=0$ can only receive $A=0$. More generally, for two treatment levels of interest $\{a',a''\}$, we may have $\{a',a''\}\subseteq \mathcal{A}_{s=1}$ but $\exists a\in \{a',a''\} ~\text{s.t.}~ a\notin \mathcal{A}_{s=0}$, where $\mathcal{A}_{s} = \text{support}(A\mid S=s)$.\footnote{Note in Section \ref{sec:identification}, we assumed $\{a',a''\}\subseteq \mathcal{A}_{s}$ for $s=0,1$.} Formally, the definition of $V$ is implied by the following restriction:
\begin{assumption}[Experimental treatment]
 $P(S=V)=1$.  
 \label{ass: exptxt}
\end{assumption}

In certain cases, defining an intervention on treatment access $V$ is a prerequisite for an intervention on $A$ to be well-defined, because accessibility (whether reflecting infrastructure, or the mode of delivery\footnote{e.g., trial-specific treatment deployment}) can influence post-treatment variables.
Crucially, considering $V$ makes it possible to identify and measure the common causes of $Y^{a,v=1}$ and $A^{v=1}$ in the target population where $A=0$ under current practice.
When the mode of delivery does not lead to differing effects, it may be reasonable to assume that context-specific accessibility affects neither the exposure nor the outcome, except indirectly through the treatment received.
\begin{assumption}[$V$-irrelevance]
\normalfont
The exclusion restriction placed on $V$ states that $V$ can only affect $M$ and $Y$ through $A$:
$M^{v=1,a} = M^a$, $Y^{v=1,a} = Y^a$.\footnote{A distributional version of exclusion restriction is given by $Y^{ v, a}, M^{v, a} \independent_d v \mid A^v, L, S$.}
    \label{ass:exclusion}
\end{assumption}
Under modified versions of the assumptions defining $\mathcal{M}^{di}$ (see Web Appendix C), we can identify $E(Y^{a,v=1}\mid S=0)$, the average counterfactual outcome had treatment been set to $a$ and had treatment $a$ been made accessible in $S=0$.
These modified assumptions mirror those in Theorem \ref{thm:ident1} but concern conditional independencies involving variables under the additional population-level intervention $v=1$. While these conditions concern isomorphic conditional independencies, they are not implied under $\mathcal{M}^{di}$ because $V=1$ represents a context not supported by the joint distributions in $S=0$ assumed under the original assumptions.
Nevertheless, the comparison between SEE and LTE follows reasoning closely paralleling Section \ref{sec: LTE-SEE} (see Web Appendix C).

\begin{theorem}
    Under Assumption \ref{ass: exptxt}, along with modified assumptions that mirror Theorem \ref{thm:ident1} given in Web Appendix C, the average counterfactual outcome had treatment been set to a value $a$ and had treatment $a$ been made accessible in $S=0$ for $a\in \mathcal{A}$, i.e., $E(Y^{a,v=1}\mid S=0)$, is identified by g-formula \eqref{eq:identeq1lte}. 
    \label{thm:ident2main}
\end{theorem}
\begin{corollary}
    By further imposing Assumption \ref{ass:exclusion}, we obtain $\Psi^a \coloneqq E(Y^{a}\mid S=0) = E(Y^{a,v=1}\mid S=0)$.
    If we assert $P(Y=Y^*)$ and  $Y\independent_d S\mid (M,L)$, then $\Psi^a$ is identified by g-formula \eqref{eq:identeq-strong}. 
\end{corollary}

 \begin{remark}
     In a similar vein as the estimand involving an intervention on $V$, we may be interested in a different estimand when Assumption~\ref{ass:treatsource1} fails in the setting of Section \ref{sec:identification}. For example, if individuals in the trial receive ancillary medical care, then it is possible that $P(M^{a=1}=m\mid L, S=1) \neq P(M^{a=1}=m \mid L,S=0).$ This suggests the presence of a variable $S_M$, representing ancillary care, that mediates the effect of $S_M$ on $M$. This is particularly relevant when, in real-world settings, implementation of $A$ will likely include these additional components as part of standard delivery, making the joint intervention on $(A,  S_M)$ a more realistic and policy-relevant target. 
 \end{remark}

\section{Efficient estimation}
\label{sec:estimation}

The efficient influence function $\varphi^{eff}({O})$ for a parameter $\Psi^a$ in a {non-parametric model $\mathcal{M}_{\text{np}}$ for the observable data is given by ${d\Psi^a(\epsilon)}/{d\epsilon}\vert_{\epsilon=0} = E\{\varphi^{eff}({O})\mathcal{S}({O})\}$, where ${d\Psi^a(\epsilon)}/{d\epsilon}\mid_{\epsilon=0}$ is a pathwise derivative of the parameter $\Psi^a$ along any parametric submodel of the observed data distribution indexed by $\epsilon$, and $\mathcal{S}({O})$ is the score function of the parametric submodel evaluated at $\epsilon=0$ \citep{newey1994,Van2000}. The efficient influence function for the functional $\Psi^a$, identified by formula \eqref{eq:identeq1lte} under sampling from $q(s,l,a,m,y)$, is given by:
\medskip
\begin{align}
     &\frac{(1-S)P(A=a\mid S=1, M,L)p(M\mid S=1,L)}{P_q(S=0)P(A=a\mid S=1,L)p(M\mid S=0,L)}\left\{Y-T_1(L,M,S) \right\} + \nonumber
     \\& \frac{S\cdot I(A=a) \cdot P_q(S=0\mid L)}{P_q(S=0)P(A=a\mid S=1, L)P_q(S=1\mid L)}\left\{T_1(L,M,S=0)-T_2(L,A,S) \right\}+ \nonumber
     \\& \frac{(1-S)}{P_q(S=0)}\left[T_2(L,A=a,S=1)-E\left\{T_2(L,A=a,S=1)\mid S\right\} \right],
     \label{eq:eif}
\end{align}

\allowdisplaybreaks\sloppy
where subscript $q$ in $P_q(\cdot)$ denotes that all quantities are evaluated under the biased sampling model; $T_1(L,M,S) \coloneqq E(Y\mid L,M, S)$, $T_2(L,A,S) \coloneqq E\left\{T_1(L,M,S=0)\mid L,A,S\right\},$ and
$\Psi^a = E\left\{T_2(L,A=a,S=1)\mid S=0\right\}$.

The first term of the efficient influence function given by \eqref{eq:eif} can be re-expressed as:
\medskip
\begin{align}
     &\frac{(1-S)}{P_q(S=0)}\frac{p(M\mid S=1, A=a,L)}{p(M\mid S=0,L)}\left\{Y-T_1(L,M,S) \right\},
     \label{eq:eif2}
\end{align}
which closely resembles a key term in the efficient influence function for corresponding stochastic exposure interventions in SEEs \citep{Wen2021}. In particular, $p(M\mid S=1, A=a,L)$ plays a role analogous to the hypothesized treatment distribution in those settings (see Section \ref{sec: LTE-SEE}).\footnote{More precisely, the efficient influence function of mean counterfactual estimand in a stochastic intervention $g$ (i.e., $E(Y^g\mid S=0)$) where exposure is taken as a random draw from a hypothesized function $p^g(M\mid L,S=0)$ is given by \citep{Wen2021}: $$\underbrace{\frac{(1-S)}{P_q(S=0)}\frac{p^g(M\mid L,S=0)}{p(M\mid L,S=0)}\left\{Y-T_1(L,M,S) \right\}}_{(\ast)} + \underbrace{\frac{(1-S)}{P_q(S=0)}\left[\sum_m T_1(L,M=m,S=0)p^g(m\mid L,S=0) - E(Y^g\mid S=0)\right]}_{(\ast\ast)},$$ where $(\ast)$ resembles \eqref{eq:eif2} and $(\ast\ast)$ resembles the third term in \eqref{eq:eif}. The second term in \eqref{eq:eif} can be viewed as an additional term for transporting the effect of $A$ on $M$ from $S=1$ to $S=0$.}
The two formulations yield distinct sets of nuisance functions to estimate. The efficient influence function in \eqref{eq:eif} requires models for $P(A=a\mid S=1,M,L)$, $P(A=a\mid S=1,L)$, $P(M=m\mid S,L)$, $P_q(S=1\mid L)$, $T_1(L,M,S)$, and $T_2(L,A,S)$\footnote{$P(A=a \mid S=1, L)=P(A=a\mid S=1)$ is known in RCTs, assuming adherence.}. In contrast, the efficient influence function based on \eqref{eq:eif2} requires $P(M=m\mid S=1, A,L)$, $P(M=m\mid S=0, L)$, $P(A=a\mid S=1,L)$, $P_q(S=1\mid L)$, $T_1(L,M,S)$, and $T_2(L,A,S)$. 
Notably, these formulations allow for the construction of triply robust estimators.
In Web Appendix F, we show that $\varphi^{eff}({O})$ is the efficient influence function under a semiparametric model $\mathcal{M}_{semi,1}$ that incorporates the restriction $Y\independent A\mid (L,M,S)$.
In Web Appendix G, we derive the efficient influence function for a treatment regime that depends on covariates $L$, motivated by the PrEP Example 2.


The structure of the efficient influence function motivates the development of new estimators $\hat\Psi^a_{TMLE}$, which solve an empirical average of the efficient influence function using data from both data sources.  
Without loss of generality, we describe a new estimator based on the efficient influence function in \eqref{eq:eif}, though it can be easily adapted to solve the efficient influence function based on \eqref{eq:eif2} by adjusting the defined weights.
In what follows, we let $\Phi^{-1}$ denote a known inverse link function satisfying $\inf(\mathbf{Y})\leq \Phi^{-1}(u)\leq \sup(\mathbf{Y})$, for all $u$, where $\mathbf{Y}$ is the sample space of $Y$ (e.g., a logit link for dichotomous $Y$).
We let $\hat e_a(L)$, $\hat f_a(M,L)$, $\hat g_M(S,L)$, and $\hat h_s(L)$ denote estimates of $P(A=a\mid S=1,L)$, $P(A=a\mid S=1, M,L)$, $ p(M\mid S, L)$ and $P_q(S=s\mid L)$, respectively.  
In the algorithm below, steps 2--3 ensure that the estimates for $T_1(L,M,S)$ and $T_2(L,A,S)$ are bounded within the parameter space with probability 1. As a result and by construction, $\hat{\Psi^a}_{TMLE}$ will also be sample-bounded (i.e., estimates from these estimators fall within the parameter space with probability 1), which is helpful with outcomes are binary, as in our data application. 
 \sloppy\raggedbottom
\subsection{A targeted maximum likelihood estimator.}
\begin{breakablealgorithm}
\renewcommand{\theenumi}{\Alph{enumi}}   
\caption{Algorithm for Targeted Maximum Likelihood Estimator (TMLE)}          
\begin{algorithmic} [1]      
\item Construct estimators for $P(A=a\mid S=1, L)$ and $P(A=a\mid S=1, M,L)$ using data from $S=1$, and $p(M\mid S,L)$ and $P_q(S=s\mid L)$ using data from both sources. Predict $P(A=a\mid S=1, L)$, $P(A=a\mid S=1, M,L)$, $p(M\mid S,L)$ and $P(S=s\mid L)$ on all subjects. 
\item 
\begin{enumerate}
    \item For those in $S=0$, obtain an initial estimate of $E(Y\mid L,M,S=0)$ denoted by $\hat T_1(L,M,S=0)$.
    \item In the same group of subjects, update the initial outcome model by fitting an intercept-only regression model $\tilde T_1(L,M,S=0;\delta_1)= \Phi^{-1}\{\Phi(\hat T_1(L,M,S=0))+\delta_1\}$ where the score function for each observation is weighted by 
    $\hat{W}_1=\frac{\hat f_a(M,L) \hat g_{M}(S=1,L) }{\hat e_a(L)\hat g_{M}(S=0,L)},$ and $\Phi(\hat T_1(L,M,S=0))$ is set as an offset term.
    \item Using the previous regression model, predict $\tilde T_1(L,M,S=0;\hat\delta_1)$ for all observations in both sets of data.
\end{enumerate}
\item 
\begin{enumerate}
    \item For those in $S=1$, obtain an initial estimate of $T_2(L,A,S=1)$ denoted by $\hat T_2(L,A,S=1)$, treating $\tilde T_1(L,M,S=0;\hat\delta_1)$ as the outcome .
    \item In the same group of subjects, update the initial outcome model by fitting an intercept-only regression model $\tilde T_2(L,A,S=1;\delta_2)= \Phi^{-1}\{\Phi(\hat T_2(L,A,S=1))+\delta_2\}$ where the score function for each observation is weighted by $\hat{W}_2=\frac{\hat h_{s=0}(L)}{ \hat e_a(L) \hat h_{s=1} (L)},$ and $\Phi(\hat T_2(L,A,S=1))$ is set as an offset term.
    \item Using the previous regression model, predict $\tilde T_2(L,A=a,S=1;\hat\delta_2)$ for all observations in both sets of data.
\end{enumerate}
\item For those in $S=0$, estimate $\Psi^a$ by the TMLE estimator denoted by $\hat{\Psi^a}_{TMLE}$ by averaging the predicted outcomes $\tilde T_2(L,A=a,S=1;\hat\delta_2)$ obtained in the previous step.
That is, let $\hat{\Psi^a}_{TMLE}=\mathbb{P}_{n_{s=0}} \{\tilde T_2(L,A=a,S=1;\hat\delta_2) \},$ where $\mathbb{P}_{n_{s=0}} (X) =  n_{s=0}^{-1}\sum_{\{\forall i: S_i=0\}} (X_i)$. 
\end{algorithmic}
\end{breakablealgorithm}

\subsubsection{Asymptotic properties.}
~Building on the notations introduced earlier, let $\hat\gamma$ denote estimate of $P_q(S=1)^{-1}>0$. Moreover, let $\gamma^\ast$, $e_a^\ast(L)$, $f_a^\ast(M,L)$, $g_M^\ast(S,L)$ and $h_s^\ast(L)$ denote their corresponding asymptotic limits. Similarly, let $T_2^\ast(L,A,S=1)$ and $T_1^\ast(L,M,S=0)$ denote the asymptotic limits of $\hat T_2(L,A,S=1)$ and $\hat T_1(L,M,S=0)$, respectively.
For general function of $\gamma'$, $e_a'(L)$ , $f_a'(M,L)$, $g_M'(S,L)$ and $ h_s'(L)$, $ T_2'(L,A,S=1)$ and $T_1'(L,M,S=0)$, define:
\begin{align*}
     H\left(\gamma', e_a', f_a', g_M', h_s', T_2',T_1' \right) = &{\gamma'}\Bigg[\frac{(1-S)f_a'(M,L)g_{M}'(1,L)}{e_a'(L)g_{M}'(0,L)}\left\{Y-T_1'(L,M,S) \right\} + \nonumber
     \\& \frac{S\cdot A \cdot h_{s=0}'(L)}{e_a'(L) h_{s=1}'(L)}\left\{T_1'(L,M,S=0)-T_2'(L,A,S) \right\}+ \nonumber
     \\& (1-S)T_2'(L,A=a,S=1)\Bigg]
\end{align*}
By construction, $\hat\Psi^a_{TMLE} = \mathbb{P}_n \left\{H\left(\hat\gamma, \hat e_a, \hat f_a, \hat g_M, \hat h_s, \hat T_2,\hat T_1 \right)\right\}$ where $\hat T_2(L,A,S=1)$ and $\hat T_1(L,M,S=0)$ are estimated via the TMLE algorithm. Under the assumptions provided below, we have the following asymptotic properties of $\hat\Psi^a_{TMLE}$:
\smallskip

\begin{theorem}
\allowdisplaybreaks
Suppose the following assumptions hold:
\begin{enumerate}[label=(\roman*)]
    \item The sequence $H\left(\hat\gamma, \hat e_a, \hat f_a, \hat g_M, \hat h_s, \hat T_2,\hat T_1 \right)$ and its limit $H\left(\gamma^\ast, e_a^\ast, f_a^\ast, g_M^\ast, h_s^\ast, T_2^\ast,T_1^\ast \right)$ fall in a Donsker class \citep{wellner2013weak};
    \item $\|H\left(\hat\gamma, \hat e_a, \hat f_a, \hat g_M, \hat h_s, \hat T_2,\hat T_1 \right) -  H\left(\gamma^\ast, e_a^\ast, f_a^\ast, g_M^\ast, h_s^\ast, T_2^\ast,T_1^\ast \right)\| \overset{a.s.}{\longrightarrow} 0;$
    \item $E\left[H\left(\gamma^\ast, e_a^\ast, f_a^\ast, g_M^\ast, h_s^\ast, T_2^\ast,T_1^\ast \right)^2 \right]<\infty$
    \sloppy\item
\begin{enumerate}[label=(\alph*)]
\item $\|T_2(L,A=a, S=1)-\hat T_2(L,A=a, S=1)\|~ \|e_a(L)-\hat e_a(L) \| = O_p(n^{-\nu})$, and
\item $\|T_2(L,A=a, S=1)-\hat T_2(L,A=a, S=1)  \|~ \| h_s(L) - \hat h_s(L) \| = O_p(n^{-\nu})$, and
\item $\|T_1(L,M,S=0) - \hat T_1(L,M,S=0)\|~ \|e_a(L)-\hat e_a(L) \| = O_p(n^{-\nu})$, and
\item $\| T_1(L,M,S=0) -\hat T_1(L,M,S=0)\|~ \| h_s(L) - \hat h_s(L) \| = O_p(n^{-\nu})$, and
\item $\| T_1(L,M,S=0) - \hat T_1(L,M,S=0)\|~ \|f_a(M,L)-\hat f_a(M,L) \| = O_p(n^{-\nu})$, and
\item $\| T_1(L,M,S=0) - \hat T_1(L,M,S=0)\|~ \|g_M(S,L)-\hat g_M(S,L) \| = O_p(n^{-\nu})$
\end{enumerate}
\end{enumerate}
for $\nu>1/2$ and where $\norm{f(x)} = \left\{\int |f(x)|^2dP(x)\right\}^{1/2}$, i.e. the $L_2(P)$ norm. 
Then $\hat\Psi^a_{TMLE}$ has the following asymptotic representation:
    $$ \sqrt{n}(\hat\Psi^a_{TMLE} - \Psi^a) \rightsquigarrow N(0,\sigma^2),~~\text{where}~\sigma^2 = \mathrm{Var}(\varphi^{\text{eff}}({O})).$$
\label{thm:asymptot}
\end{theorem}

\begin{corollary}
The nonparametric efficiency bound under biased sampling is given by the variance of the efficient influence function which can be expressed by: 

\begin{align*}B_q =& \frac{1}{P_q(S=0)}E\left[\frac{P(M\mid S=1,A=a,L)^2}{P(M\mid S=0,L)^2}\sigma_Y^2(L,M) + \sigma_{T_2}^2 \Big\vert S=0\right] + 
\\&\frac{1}{P_q(S=0)}E\left[\frac{P_q(S=0\mid L)^2}{P_q(S=1\mid L)^2P(A=a\mid S=1,L)}\sigma_{T_1}^2(L) \Big\vert S=1\right]
\end{align*}

where $\sigma_Y^2(L,M) = \mathrm{Var}(Y\mid L,M,S=0)$, $\sigma_{T_2}^2 = \mathrm{Var}(T_2(L,A=a,S=1)\mid S=0)$ and $\sigma_{T_1}^2(L) =\mathrm{Var}(T_1(L,M,S=0)\mid L,A=a,S=1)$.
Compared with the efficiency bound for an counterfactual mean outcome under a stochastic intervention in which values of $M$ are drawn from a pre-specified distribution $p^g(m\mid l)$ in $S=0$, the efficiency bound for $E(Y^{a}\mid S=0)$ is greater by ${P_q(S=0)}^{-1}E\left[\frac{P_q(S=0\mid L)^2}{P_q(S=1\mid L)^2P(A=a\mid S=1,L)}\sigma_{T_1}^2(L) \Big\vert S=1\right].$ 
\end{corollary}

This represents an important bias-variance trade-off: imposing a function $p^g(m\mid l)$ for $P(M^a=m\mid L=l, S=0)$ can reduce variance, but may introduce bias if the assumed relationship is incorrect. In practice, such a function is unlikely to be correctly specified, making it hard to define a valid $p^g(\cdot)$ without strong, often unverifiable, assumptions. In Web Appendix F, we compare against stochastic interventions in which the \textit{hypothesized} distribution of $M$ depends on the observed distribution of $M$ in $S=0$.

\begin{corollary}
Under standard regularity conditions and working models for the nuisance functions, $\hat\Psi^a_{TMLE}$ will be consistent and asymptotically normal under the union model $\mathcal{M}_{union}=\mathcal{M}_{1}\cup \mathcal{M}_2\cup \mathcal{M}_3$ where we define:
        \begin{enumerate}[label=(\alph*)]
        \item Model $\mathcal{M}_1$: $\hat e_a(L) \overset{a.s.}{\longrightarrow} e_a^\ast(L) = P(A=a\mid S=1, L)>\epsilon$ for some $\epsilon>0$\footnote{Holds in RCTs, assuming adherence.}, \\
        $\hat f_a(M,L) \overset{a.s.}{\longrightarrow} f_a^\ast(M,L) = P(A=a\mid S=1, M, L)>\epsilon$ for some $\epsilon>0$, 
        \\$\hat g_M(S,L) \overset{a.s.}{\longrightarrow} g_M^\ast(S,L) = p(M\mid S,L)>\epsilon$ for some $\epsilon>0$, and 
        \\$\hat h_s(L) \overset{a.s.}{\longrightarrow} h_s^\ast(L) = P_q(S=s\mid L)>\epsilon$ for some $\epsilon>0$; \textbf{or}
        \item Model $\mathcal{M}_2$: $\hat T_1(L,M,S=0)\overset{a.s.}{\longrightarrow} T_1^\ast(L,M,S=0) =T_1(L,M,S=0)$ and 
        \\$\hat T_2(L,A=a,S=1)\overset{a.s.}{\longrightarrow} T_2^\ast(L,A=a,S=1) = T_2(L,A=a,S=1)$; \textbf{or}
        \item Model $\mathcal{M}_3$: $\hat T_1(L,M,S=0)\overset{a.s.}{\longrightarrow} T_1^\ast(L,M,S=0) =T_1(L,M,S=0)$,
        \\ $\hat e_a(L) \overset{a.s.}{\longrightarrow} e_a^\ast(L) = P(A=a\mid S=1, L)>\epsilon$ for some $\epsilon>0$, and
        \\$\hat h_s(L) \overset{a.s.}{\longrightarrow} h_s^\ast(L) = P_q(S=s\mid L)>\epsilon$ for some $\epsilon>0$.
    \end{enumerate}
\end{corollary}

 When the nuisance functions are estimated with machine learning algorithms, the variance of the TMLE can be estimated empirically with $\mathbb{P}_n[\left\{\hat\varphi^{\text{eff}}({O})\right\}^2]$, where all nuisance functions are replaced with their estimates. When the nuisance functions are estimated using parametric models, this variance estimator remains valid as long as all nuisance functions are correctly specified. However, in practice, we recommend using the non-parametric bootstrap to estimate the variance, as parametric models are more prone to misspecification.

 \subsection{Special case: Outcome is measured in $S=1$.}
 \label{sec:special}
~The efficient influence function for $\Psi^a = E(Y^a\mid S=0)$ under sampling from $q(s,l,a,m,y)$ and under a semiparametric model $\mathcal{M}_{semi,2}$ that satisfies $Y\independent (A,S)\mid (L,M)$ is given by:

 \begin{align}
   \varphi^{eff}({O})  &=\frac{P(A=a\mid S=1, M,L)p(M\mid L,S=1)P_q(S=0\mid L)}{P(A=a\mid S=1, L)\sum_s p(M\mid L,S=s)P_q(S=s\mid L)P_q(S=0)}\left\{Y-T_1(L,M) \right\} + \nonumber
     \\& \frac{S\cdot A \cdot P_q(S=0\mid L)}{P(A=a\mid S=1, L)P_q(S=1\mid L)P_q(S=0)}\left\{T_1(L,M)-T_2(L,A,S) \right\}+ \nonumber
     \\& \frac{(1-S)}{P_q(S=0)}\left[T_2(L,A=a,S=1)-E\left\{T_2(L,A=a,S=1)\mid S\right\} \right]
     \label{eq:eifspecial}
 \end{align}
 \sloppy
 
where we define $T_1(L,M) \coloneqq E(Y\mid L,M)$ and $T_2(L,A,S) \coloneqq E\left\{T_1(L,M)\mid L,A,S\right\}$. 
The first term of the efficient influence function \eqref{eq:eifspecial} admits an alternative representation given by $\frac{P_q(S=0\mid L)}{P_q(S=0)}\frac{p(M\mid S=1, A=a,L)}{p(M\mid L)}\left\{Y-T_1(L,M) \right\}$, illustrating once again a connection to stochastic exposure interventions.
The TMLE estimator based on \eqref{eq:eifspecial} follows analogously to Algorithm 1 and is provided in Web Appendix F, along with its asymptotic properties which parallel those presented in Theorem \ref{thm:asymptot}.
Estimators based on \eqref{eq:eifspecial} allows one to model the outcome by pooling over different data sources, thereby improving precision in estimation. 
Since the semiparametric model $\mathcal{M}_{semi,2}$ that satisfies $Y\independent (A,S)\mid (L,M)$ is necessarily nested in the semiparametric model $\mathcal{M}_{semi,1}$ that satisfies $Y\independent A\mid (L,M)$. As such, the estimators based on the efficient influence function in $\mathcal{M}_{semi,2}$ will be at least as efficient as estimators based on the efficient influence function in $\mathcal{M}_{semi,1}$ (\citealp{Tsiatis2006}).

\section{Simulation study}
\label{sec:simulations}
We begin by simulating cohorts of $100,000$ individuals.
In each cohort we vary the number of randomized participants, $n_{s=1}$, across six scenarios $(250,~500,~750,~1000,~1500,~2000)$, and and set the number of non-randomized individuals $n_{s=0}$ to one of $(2500,~3000,~5000)$.
Following \citet{dahabreh2023efficient}, the data generation process was conducted as follows: generation of covariates from a cohort which include all individuals from the observational study and the RCT; selection for trial participation who are not in the observational study; simple random sampling of individuals from the remaining non-randomized individuals, with the remaining non-randomized individuals constituting the target population; (random) treatment assignment; post-treatment exposure $M$ and outcome of interest $Y$ generation. 
Specifically:
\begin{enumerate}
    \item \textit{Covariates}: simulate measured covariates from a cohort. In this simulation, we generate $L=(L_1,L_2,L_3) $ where $L_1\sim N(0,1),~L_2\sim N(0,1)$ and $L_3\sim \text{Unif}(0,2).$
    \item \textit{Selection for trial participation:} We ``select'' observations for trial participation using a logistic-linear model such that the probability of being selected into the trial is given by $P(S=1\mid L) = \text{expit}(\beta_0+L_1+L_2+L_3-0.5L_3^2)$, where the intercept $\beta_0$ is determined numerically to achieve the desired average sample size $n_{s=1}=(250,~500,~750,~1000,~1500,~2000)$. 
    \item \textit{Sampling of individuals into observational study:} We took a simple random sample of $n_{s=0}$ of non-participants in the simulated cohort of sample size $n$. 
    \item \textit{Treatment in observational study:} Treatment variable $A$ is set to 0 for all individuals in the observational study (i.e., $A=1$ is not yet available in $S=0$). 
    \item \textit{Treatment in RCT:} We generate an indicator of unconditionally randomized $A$ assignment among randomized individuals. The $A$ assignment mechanism is given by $A\sim \text{Ber}(P(A=1\mid S=1))$ where $P(A=1\mid S=1)=0.5$.
    \item \textit{Exposure in observational study and the RCT:} We generate exposure in both studies using $M\sim \text{Ber}(P(M=1\mid A, L))$ where $P(M=1\mid A,L)=\text{expit}(-3+2A+2L_1+L_2+L_3)$.
    \item \textit{Outcomes in observational study:} We generate outcomes in $S=0$ using $Y\sim \text{Ber}(P(Y=1\mid M,L,S)$ where $P(Y=1\mid M,L,S) = \text{expit}(1+2M+L_1+L_2-L_3-L_2\cdot L_3- L_2^2+S)$.
\end{enumerate}
The simulation study aims to compare the performance of TMLE with singly robust estimators -- iterative conditional expectation (ICE) and inverse probability weighted (IPW) estimators -- when the nuisance functions are estimated through machine learning algorithms (see Web Appendix D for more details for these singly robust estimators). 
We used  generalized additive models to estimate all of the nuisance functions. Under certain smoothness conditions (e.g., twice continuously differentiable additive components; \citealp{horowitz2004nonparametric,horowitz2009semiparametric}), generalized additive model estimators can achieve the $n^{2/5}$ rate of convergence. Additional simulation details are provided in Web Appendix H. 

Figure~\ref{fig:lab2}  and Tables~\ref{tab:results}--\ref{tab:MLCP} compare the performance of the three estimators based on 1000 simulated datasets for $E(Y^{a=1,v=1}\mid S=0)=E(Y^{a=1}\mid S=0)$. The scaled bias (scaled by $\sqrt{n}$) of single robust estimators are greater than those of TMLE in all scenarios, with greater discrepancy as $n_{s=0}$ increases. 
This is expected because the ICE and IPW estimators are not expected to converge at $\sqrt{n}$ rates when machine learning is used for nuisance parameter estimation, whereas TMLE allows the nuisance functions to converge at slower nonparametric rates. Thus, the bias of TMLE tends to zero faster compared with the other estimators, albeit with a slightly larger standard error compared with ICE at smaller sample sizes. 
Moreover, the estimated coverage probability of the confidence intervals for TMLE based on the asymptotic variance gets closer to the nominal 95\% as sample size of $S=1$ increases. For instance, for $n_{s=0}=3000$, the 95\% coverage probability is $(81.6,~89.1,~90.7,~92.1,~94.5,~94.7)$ for $n_{s=1}= (250,~500,~750,~1000,~1500,~2000)$, respectively.

Additional simulation studies conducted using parametric nuisance functions are provided in Web Appendix H. 
Specifically, we consider a simulation setting similar to the one described in the main text, but with the nuisance functions estimated using parametric models. We also consider a scenario where treatment level in $S=0$ is fixed at $A=0$ for all individuals, while treatment levels $A=\{1,2\}$ are accessible in $S=1$, mimicking scenarios where novel treatment methods being compared are not yet available in real world settings. 
These supplementary simulation studies demonstrate the identification of our estimand in these specific contexts, as well as the model robustness property of the TMLE estimator.

In Web Appendix H, we further evaluate the performance of our TMLE estimators under the outcome-source conditional independence assumption, specifically in settings where outcomes $Y$ are observed for all individuals in $S=1$. The results indicate that, when $Y\independent_d S\mid (M,L)$ holds, the TMLE estimator for $\Psi^a$ that allows for pooling outcome data across sources is more efficient than the estimator that does not permit pooling. This gain in efficiency is expected, as the pooling estimator leverages the additional assumption of outcome-source conditional independence to incorporate outcome information from all sources when estimating the causal estimand.

\begin{figure}
\centering
\includegraphics[width=1\textwidth]{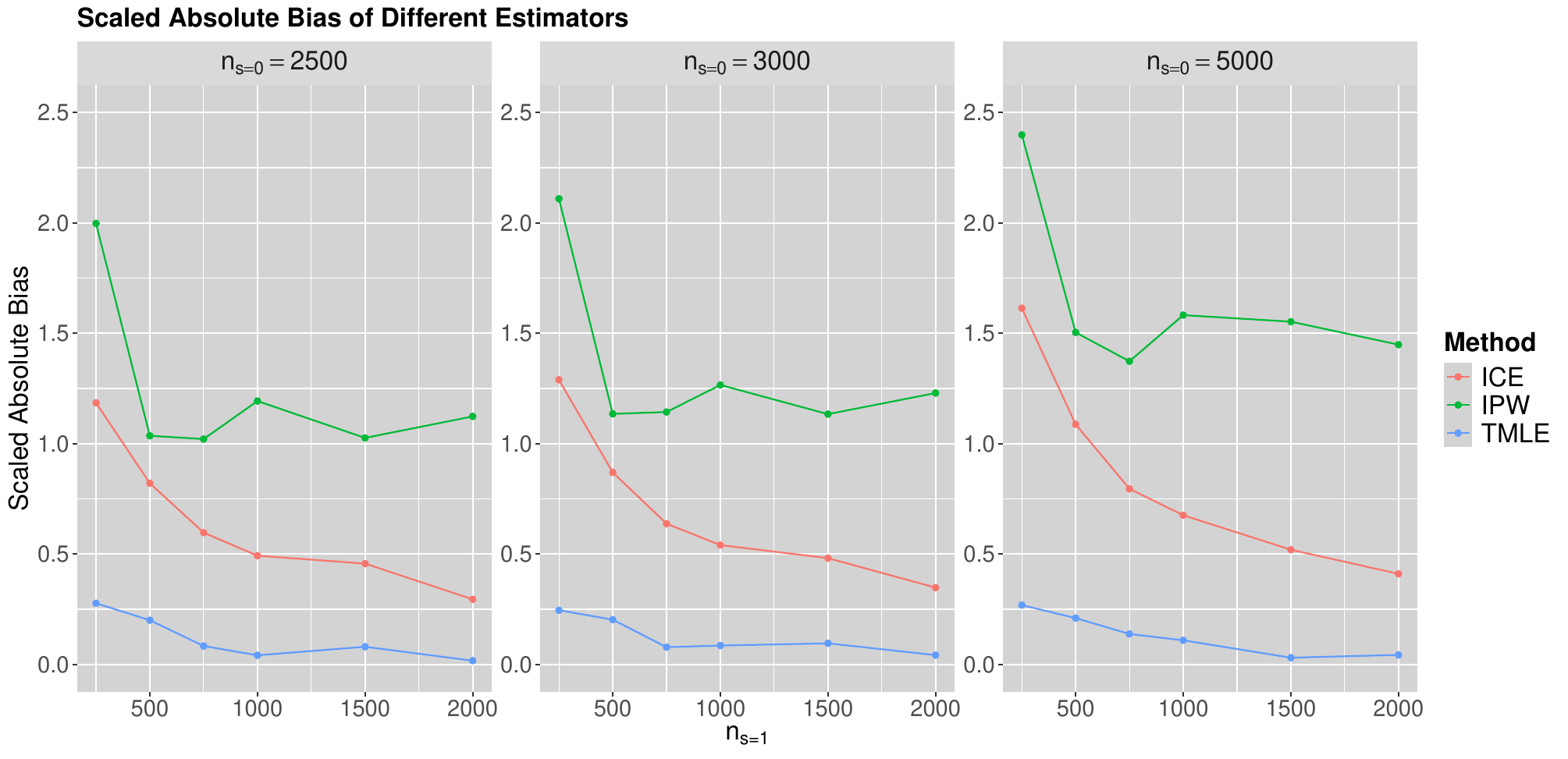}
 \caption{Scaled absolute bias (scaled by $\sqrt{n}$) for simulation study. TMLE denotes the the TMLE estimator; ICE denotes singly robust Iterative Conditional Expectation estimator; IPW denotes singly robust Inverse Probability Weighted estimator.}
  \label{fig:lab2}
\end{figure}

\renewcommand{\arraystretch}{1.15}
\begin{table}
\centering
\begin{tabular}{|c|c|c|c|c|c|c|c|c|c|}
\hline
\multicolumn{10}{|c|}{$n_{s=0} = 2500$} \\
\hline
& \multicolumn{3}{|c|}{Bias} & \multicolumn{3}{|c|}{SE} & \multicolumn{3}{|c|}{MSE}  \\
\hline
& \textbf{TMLE} & \textbf{ICE} & \textbf{IPW} & \textbf{TMLE} & \textbf{ICE} & \textbf{IPW} & \textbf{TMLE} & \textbf{ICE} & \textbf{IPW} \\
\hline
$n = 250$  & 0.528  & 2.259  & -3.809  & 6.152  & 3.749  & 8.916  & 6.175  & 4.377  & 9.696  \\
$n = 500$  & 0.366  & 1.498  & -1.891  & 2.484  & 2.520  & 4.426  & 2.511  & 2.932  & 4.813  \\
$n = 750$  & 0.146  & 1.047  & 1.79    & 2.550  & 2.188  & 3.799  & 2.554  & 2.426  & 4.200  \\
$n = 1000$ & 0.070  & 0.832  & -2.017  & 2.534  & 1.943  & 3.469  & 2.535  & 2.114  & 4.013  \\
$n = 1500$ & 0.126  & 0.722  & -1.622  & 2.083  & 1.690  & 3.032  & 2.087  & 1.838  & 3.439  \\
$n = 2000$ & 0.025  & 0.440  & -1.675  & 1.456  & 1.519  & 2.389  & 1.456  & 1.581  & 2.917  \\
\hline
\multicolumn{10}{|c|}{$n_{s=0} = 3000$} \\
\hline
$n_{s=1} = 250$  & 0.431  & 2.262  & -3.700  & 6.505  & 3.699  & 8.843  & 6.519  & 4.335  & 9.586  \\
$n_{s=1} = 500$  & 0.343  & 1.470  & -1.918  & 2.456  & 2.458  & 4.503  & 2.480  & 2.864  & 4.894  \\
$n_{s=1} = 750 $ & 0.128  & 1.041  & -1.867  & 2.514  & 2.079  & 4.025  & 2.517  & 2.325  & 4.436  \\
$n_{s=1} = 1000$ & 0.135  & 0.855  & -2.001  & 1.893  & 1.886  & 3.084  & 1.898  & 2.071  & 3.677  \\
$n_{s=1} = 1500$ & 0.142  & 0.717  & -1.690  & 1.552  & 1.600  & 3.005  & 1.558  & 1.753  & 3.447  \\
$n_{s=1} = 2000$ & 0.059  & 0.492  & -1.738  & 1.399  & 1.442  & 2.676  & 1.400  & 1.524  & 3.191  \\
\hline
\multicolumn{10}{|c|}{$n_{s=0} = 5000$} \\
\hline
$n_{s=1} = 250$  & 0.371  & 2.227  & -3.311  & 6.678  & 3.604  & 8.735  & 6.689  & 4.236  & 9.341  \\
$n_{s=1} = 500 $ & 0.283  & 1.466  & -2.028  & 2.763  & 2.333  & 5.098  & 2.777  & 2.756  & 5.486  \\
$n_{s=1} = 750$  & 0.182  & 1.049  & -1.811  & 1.958  & 1.941  & 3.204  & 1.967  & 2.206  & 3.680  \\
$n_{s=1} = 1000$ & 0.141  & 0.872  & -2.042  & 1.722  & 1.710  & 3.070  & 1.728  & 1.919  & 3.687  \\
$n_{s=1} = 1500$ & 0.038  & 0.644  & -1.926  & 2.129  & 1.481  & 3.135  & 2.130  & 1.615  & 3.679  \\
$n_{s=1} = 2000$ & 0.051  & 0.490  & -1.731  & 1.232  & 1.278  & 2.198  & 1.233  & 1.369  & 2.798  \\
\hline
\end{tabular}
\caption{Results for simulation study for various sample sizes: $n_{s=1}=(250,~500,~750,~1000,~1500,~2000)$ and $n_{s=0}=(2500,~3000,~5000)$ using nonparametric models for nuisance functions: Bias, standard error (SE), and mean squared error (MSE) are multiplied by 100. }
\label{tab:results}
\end{table}

\begin{table}
\centering
\begin{tabular}{|c|c|c|c|}
\hline
 & $n_{s=0} = 2500$ & $n_{s=0} = 3000$ & $n_{s=0} = 5000$ \\
\hline
$n_{s=1} = 250$ & 82.2 & 81.6 & 79.4 \\
$n_{s=1} = 500$ & 90.5 & 89.1 & 88.2 \\
$n_{s=1} = 750$ & 91.1 & 90.7 & 91.2 \\
$n_{s=1} = 1000$ & 93.1 & 92.1 & 92.0 \\
$n_{s=1} = 1500$ & 94.5 & 94.5 & 93.6 \\
$n_{s=1} = 2000$ & 96.2 & 94.7 & 94.8 \\
\hline
\end{tabular}
\caption{Coverage of proposed TMLE estimator at varying sample sizes of $S=0$ and $S=1$.}
\label{tab:MLCP}
\end{table}

\section{Data analysis}
\label{sec:datanalysis}
 For illustrative purposes, we utilize trial data from the Clinical Trials Network, a publicly accessible repository of substance use-related randomized controlled trials funded by the National Institute on Drug Abuse. The trial of interest, CSP-1022, is a double-blind, placebo-controlled trial conducted to evaluate the efficacy and safety of the selegiline transdermal system (STS) combined with cognitive behavioral intervention for smoking cessation in heavy smokers aged 18 or older who are generally in good health \citep{kahn2012selegiline}.
A total of 246 men and women were randomly assigned to receive either the STS patch  and behavioral intervention ($n = 121$) or a placebo patch and behavioral intervention ($n = 125$) for a duration of 8 weeks. Our exposure (mediating exposure) of interest is whether an individual is still smoking at the 6-month follow-up from treatment randomization.

We integrate the data from the CSP-1022 trial with a representative sample from a target population consisting of all U.S. adults (18 years or older) who are current smokers at baseline and are generally in good health. To obtain a reasonable sample from this target population, we used data from the 2013–2014 National Health and Nutrition Examination Survey (NHANES). NHANES is a cross-sectional survey designed to be representative of the non-institutionalized U.S. population, collecting detailed health and lifestyle information through interviews, physical examinations, and laboratory tests. By leveraging NHANES data, we aim to evaluate the effect of the interventions from the CSP-1022 trial on long-term outcomes in a broader population. Specifically, we are interested in estimating the effect of assigned STS and behavioral intervention on 5-year all-cause mortality in the NHANES study. 
In both studies, individuals' smoking status is assessed at the 6-month mark (which corresponds to the survey time for NHANES). In NHANES, all individuals aged 18 and older who have ever smoked and reported smoking six months before the time of survey are included in the dataset ($n_{s=0}=1319)$. We assume that NHANES participants did not have access to the STS or the specific smoking-related cognitive behavioral interventions that were offered in the trial.

Data from the CSP-1022 trial include information on individuals' treatment intervention ($A$, where $A=1$ indicates the STS intervention and $A=2$ indicates placebo), post-treatment smoking status ($M$, with $M=1$ indicating the individual is still smoking and $M=0$ otherwise), and covariates ($L$). These covariates include age, sex assigned at birth (male or female), race (non-Hispanic White, non-Hispanic Black, or other), education level (high school or less, or more than high school), body mass index (BMI), marital status, and whether the individual has been a long-term smoker (defined as smoking for more than 20 years).
Data from NHANES include information on individuals' smoking status ($M$), five-year all-cause mortality ($Y$, where $Y=1$ if subject died and 0 otherwise), and covariates ($L$) as listed above. We considered the counterfactual cumulative incidence $E(Y^{a,v=1}\mid S=0)$ ($a=1,2$) and the causal contrast given by $E(Y^{a=1,v=1}\mid S=0) - E(Y^{a=2,v=1}\mid S=0)$. 
Here, $E(Y^{a=1,v=1}\mid S=0)$ is the cumulative risk in NHANES if individuals received the combined STS plus behavioral intervention with accessibility defined by in-person delivery as in the trial, and $E(Y^{a=2,v=1}\mid S=0)$ is the cumulative risk if individuals received the placebo patch plus behavioral intervention under the same accessibility regime. These estimands were estimated using the TMLE estimator by specifying logistic regression models for the outcome, exposure, treatment and data-source. All models included linear main effects for the baseline covariates and their pairwise interactions as listed above. 

The original trial found that the STS intervention group showed a slightly higher rate of sustained smoking cessation compared to the placebo group \citep{kahn2012selegiline}. However, the effect size was not statistically significant at the end of treatment or during any follow-up assessments, likely because of the limited sample size in the randomized trial.
When the effect of smoking cessation intervention was transported to the NHANES sample, the estimated risk difference was $-0.136\%$ (95\% CI:~$[-0.983\%,~0.710\%]$ via 500 non-parametric bootstrap samples), which suggest no significant difference in five-year cumulative risk between the STS intervention arm and the placebo arm in the NHANES sample. 

Two additional sensitivity analyses were conducted to assess the robustness of our findings. The first sensitivity analysis was conducted under the assumption that smoking status remains unchanged from six months to one year after randomization in the CSP-1022 trial. Consequently, our sensitivity analysis incorporated NHANES participants who were smoking one year prior to the time of survey. The findings of this sensitivity analysis aligned closely with those of the primary analysis. In particular, the estimated causal effect was $-0.221\%$ (95\% CI: [-1.430\%, 0.989\%]). The second sensitivity analysis included an additional exposure variable, namely whether the individual was still smoking at the three-month follow-up after treatment randomization. With both post-treatment exposure variables included, the estimated causal effect was $-0.621\%$ (95\% CI: [-2.862\%, 1.619\%]), suggesting that the primary findings are generally robust to the inclusion of additional mediating factors related to smoking behavior.

\section{Discussion}

We have outlined methods for estimating LTEs in a target population when (i) the treatment under study may be available but unmeasured, or (ii) one or more treatment levels are entirely unavailable.
Unlike existing stochastic exposure interventions that specify a hypothetical distribution for a post-treatment variable, we take a more principled approach by leveraging data from another study and using the observed distribution of that post-treatment variable from that source.
We developed a state-of-the-art TMLE algorithm for our transportability causal estimand, accommodating scenarios where treatment access is limited to trial settings. 
This is particularly relevant in cases where treatments are still in the experimental phase and have not been widely adopted outside of trial settings. 
Our work highlights the importance of developing statistical tools that can adapt to the complexities of transporting trial findings to real-world settings, with the ultimate goal of informing decision-making in both research and practice.


A potential limitation of our findings is that the assumption that treatment $A$ influences outcome $Y$ only through exposure $M$ may be restrictive. This is particularly true in scenarios where $M$ varies over time, and $A$ has the potential to affect future values of $M$ beyond the trial's follow-up period. We explore this issue in greater detail in Web Appendix G. However, within the context of our data analysis, this assumption may be plausible. Short-term smoking interventions, such as those examined in our study, are generally expected to effect short-term changes in smoking behavior (e.g., within the first few months), rather than exerting long-term effects. 

Another limitation in our study arises from the differences in the populations studied in the CSP-1022 Trial and the NHANES study. The CSP-1022 Trial specifically targeted heavy smokers, while the NHANES study included a broader sample of all current smokers. Our analysis assumes that, after adjusting for baseline covariates, the probability of individuals smoking under the STS intervention is the same for heavy smokers as it is for current ever smokers. This assumption is strong because heavy smokers may have different behavioral patterns, levels of addiction, or responses to interventions compared to the general population of current smokers (i.e., there may be an arrow from $S\rightarrow M$). 
Despite this, this assumption allows us to generalize the findings from the CSP-1022 Trial to a wider population, which is necessary for broader policy and intervention implications. Future work will look into sensitivity analyses on the violation of this assumption to ensure the robustness of the conclusions.

\section*{ACKNOWLEDGMENTS}
The authors extend their gratitude to Dr. Issa Dahabreh and Dr. Jon Steingrimsson for insightful discussions on the manuscript. Lan Wen is supported by the Natural Sciences and Engineering Research Council of Canada (NSERC) Discovery Grant [RGPIN-2023-03641, DGECR-2023-00455]. 
The information reported here results from secondary analyses of data from clinical trials conducted by the National Institute on Drug Abuse (NIDA). Specifically, data from NIDA-CSP-1022 were included. NIDA databases and information are available at http://datashare.nida.nih.gov.


\newpage
\bibliographystyle{apalike}
\bibliography{refs}

\end{document}